\documentclass[
prd,%
aps,%
final,%
notitlepage,%
oneside,%
onecolumn,%
nobibnotes,%
nofootinbib,%
superscriptaddress,%
noshowpacs,%
english]%
{revtex4-1}

\usepackage[T2A]{fontenc}
\usepackage[cp866]{inputenc}
\usepackage{textcomp}
\usepackage{graphicx}
\usepackage{esint}

\usepackage{babel}

\newcommand{\mDiqCC}{3.13}
\newcommand{\RDiqCC}{0.523}
\newcommand{\RMesonCC}{0.75}
\newcommand{\kappaDiqCC}{12.8}
\newcommand{\kappaMesonCC}{42.8}
\newcommand{\MtetraCC}{6.124}
\newcommand{\MScalarCC}{5.966}
\newcommand{\DMScalarCC}{-228.}
\newcommand{\MAxialCC}{6.051}
\newcommand{\DMAxialCC}{-142.}
\newcommand{\MTensorCC}{6.223}
\newcommand{\DMTensorCC}{29.5}

\newcommand{\mDiqBB}{9.72}
\newcommand{\RDiqBB}{1.35}
\newcommand{\RMesonBB}{2.27}
\newcommand{\kappaDiqBB}{5.52}
\newcommand{\kappaMesonBB}{27.1}
\newcommand{\MtetraBB}{18.857}
\newcommand{\MScalarBB}{18.754}
\newcommand{\DMScalarBB}{-544.}
\newcommand{\MAxialBB}{18.808}
\newcommand{\DMAxialBB}{-490.}
\newcommand{\MTensorBB}{18.916}
\newcommand{\DMTensorBB}{-382.}

\newcommand{\mDiqBC}{6.45}
\newcommand{\RDiqBC}{0.722}
\newcommand{\RMesonBC}{1.29}
\newcommand{\kappaDiqBC}{6.43}
\newcommand{\kappaMesonBC}{27.1}
\newcommand{\MtetraBC}{12.491}
\newcommand{\MScalarBCa}{12.359}
\newcommand{\DMScalarBCa}{-191.}
\newcommand{\MScalarBCb}{12.471}
\newcommand{\DMScalarBCb}{-78.7}
\newcommand{\MAxialBCa}{12.424}
\newcommand{\DMAxialBCa}{-126.}
\newcommand{\MAxialBCb}{12.488}
\newcommand{\DMAxialBCb}{-62.5}
\newcommand{\MAxialBCc}{12.485}
\newcommand{\DMAxialBCc}{-64.9}
\newcommand{\MTensorBC}{12.566}
\newcommand{\DMTensorBC}{16.1}

\newcommand{\bQ}{{\bar Q}}
\newcommand{\Dq}{\mathcal{D}}
\newcommand{\bDq}{{\bar\Dq}}
\newcommand{\GeV}{\mathrm{GeV}}
\newcommand{\MeV}{\mathrm{MeV}}
\newcommand{\pb}{\mathrm{pb}}
\newcommand{\fb}{\mathrm{fb}}

\makeatother

\begin{document}

\title{Tetraquarks composed of 4 heavy quarks}

\author{A.V. Berezhnoy}
\email{Alexander.Berezhnoy@cern.ch}
\affiliation{SINP of Moscow State University, Russia}

\author{A.V. Luchinsky}
\email{Alexey.Luchinsky@ihep.ru}
\affiliation{Institute for High Energy Physics, Protvino, Russia}

\author{A.A. Novoselov}
\email{Alexey.Novoselov@cern.ch}
\affiliation{Institute for High Energy Physics, Protvino, Russia}

\begin{abstract}
In the current work spectroscopy and possibility of observation at the LHC
of tetraquarks composed of $4$ heavy quarks is discussed. Tetraquarks concerned are
$T_{4c}=[cc][\bar{c}\bar{c}]$, $T_{4b}=[bb][\bar{b}\bar{b}]$ and $T_{2[bc]}=[bc][\bar{b}\bar{c]}$.
By solving nonrelativistic Schroedinger equation masses of these states are found with the hyperfine
splitting accounted for. It is shown that masses of tensor tetraquarks $T_{4c}(2^{++})$ and $T_{2[bc]}(2^{++})$
are high enough to observe these states as peaks in the invariant mass distributions of heavy quarkonia
pairs in $pp\to T_{4c}+X\to2J/\psi+X$,
$pp\to T_{2[bc]}+X\to2B_{c}+X$ and $pp\to T_{2[bc]}+X\to J/\psi\Upsilon(1S)+X$ channels while
$T_{4b}$ is under the threshold of decay into a vector bottomonia pair.
\end{abstract}
\maketitle

\section{Introduction}

Recent observation of $J/\psi$-meson pairs production in proton-proton
collisions at $7~\mathrm{TeV}$ energy at LHC renewed interest to the $4$
heavy quarks final states. In the low invariant mass region these quarks
can form bound states (called tetraquarks) which can be produced in hadronic
experiments. Thereby we would like to discuss physics of these states,
elaborate their mass spectrum and possibility of experimental observation.

Conception of tetraquarks, i.e. mesons composed of $4$ valent quarks ($qq\bar{q}\bar{q}$),
was first introduced in works \cite{Jaffe:1976ig,Jaffe:1976ih} in 1976. For example,
$a_{0}$-meson and $\sigma$-mesons was treated as possible tetraquark candidate \cite{Maiani:2004vq,Gershtein:2006ng,Mathur:2006bs,Prelovsek:2009bk}.
However, it is hard to determine quark composition of a particle in the light meson domain so
these ideas were not developed further. Observation of new unexpected states such as $X(3872)$ \cite{Abazov:2004kp,Aubert:2004ns}
gave this idea a new impetus \cite{Swanson:2006st,Ebert:2005nc}. Eccentricity of these particles consists in fact that according to the
modes of their production and decay they contain a $c\bar{c}$ pair but they can not be included
in the well known systematics of charmonia. Later similar particles were also found in the bottomonia sector \cite{Abe:2007tk,Adachi:2008pu,Ali:2009es,Ali:2011ug}.
It is natural to ascribe these mesons to tetraquarks $(Qq\bar{Q}\bar{q})$, where $q$ are light
and $Q$ --- heavy ($b$ or $c$) quarks.

However, situation when all quarks composing a tetraquark are heavy was not treated in details yet.
This possibility seems to be quite interesting as in this case determination of meson`s quark composition
becomes simpler and its parameters can be determined by solving
nonrelativistic Schroedinger equation. Our work is devoted to these particular questions.

In our recent paper \cite{Berezhnoy:2011xy} we consider tetraquark $T_{4c}=[cc][\bar c\bar c]$ in the framework of diquark model.The hyperfine splitting in that paper was described through interaction of total diquark spins. Now we would like to study also tetraquarks $T_{4b}=[bb][\bar b\bar b]$ and $T_{2[bc]}=[bc][\bar{b}\bar{c}]$. The last state is especially interesting since, in contrast to tetraquarks build from four identical quarks, both singlet and triplet spin states of the diquark are possible. It is clear, that hyperfine interaction of spin-singlet diquark cannot be described with the method used in our previous paper, so some other approach should be applied.

In the following section spectroscopy of $(cc\bar{c}\bar{c})$,
$(bb\bar{b}\bar{b})$ and $(bc\bar{b}\bar{c})$ tetraquarks is concerned.
Possibility of observation of these particles in hadronic experiments is discussed in the third section. 
We summarize our results in the short conclusion.

\section{Spectroscopy}

\subsection{General preliminaries}

In the current study diquark model of tetraquark is used. According to this approach tetraquark
\[
T=Q_{1}Q_{2}\bQ_{3}\bQ_{4}
\]
consists of 2 almost point-like diquarks $\bDq_{12}=[Q_{1}Q_{2}]$
and $\Dq_{34}=[\bQ_{3}\bQ_{4}]$ with certain quantum numbers
(such as angular momentum, spin, color) and mass.
What concerns color configuration, two quarks in diquark can be in triplet or
sextet color state. According to the manuscript \cite{Terasaki:2010mk} in the sextet
configuration diquarks experience mutual repulsion so we restrict ourselves to
the (anti)triplet color configurations. 
Angular momentum of the diquark system equals $0$ in the ground state, so 
its spin is equal to the sum of quark spins which is $0$ or $1$.
%
What concerns diquark mass, it can be determined by solving Schroedinger equation with a
correctly selected potential. According to work \cite{Kiselev:2002iy} it is possible to use
quark-antiquark interaction potential used in heavy quarkonia calculations with additional factor $1/2$ due to
the different color structures. 

In the diquark model tetraquark mass can be
determined by solving $2$-particle Schroedinger equation with point-like diquarks. 
As $[Q_{1}Q_{2}]$ and
$[\bQ_{3}\bQ_{4}]$ diquarks are in (anti)triplet color configuration, potential
of their interaction coincides with that of quark and antiquark in heavy quarkonia.
Hyperfine splitting in this system can be described by the hamiltonian \cite{Maiani:2004vq}
\begin{eqnarray}
H & = & M_{0}+2\sum_{i<j}\kappa_{ij}\left(\mathbf{S}_{i}\mathbf{S}_{j}\right),\label{eq:H}
\end{eqnarray}
where $M_{0}$ is the tetraquark mass without splitting,
$\mathbf{S}_{i}$ --- spin operator of $i$-th (anti)quark, and $\kappa_{ij}$
are constants determined from experimental data analysis or theoretically.
When dealing with potential models, $\kappa_{ij}$ coefficient can be obtained from 
the value of the $Q_{i}Q_{j}$ system wave function at origin:
\begin{eqnarray}
\kappa_{ij} & = & \frac{1}{2}\frac{4}{9m_{Q_{1}}m_{Q_{2}}}\alpha_{s}\left|R_{[ij]}(0)\right|^{2}\label{eq:kappaMeson}
\end{eqnarray}
when both $Q_{i}$ and $Q_{j}$ are quarks or antiquarks in the color triplet state, and
\begin{eqnarray}
\kappa_{ij} & = & \frac{1}{2}\frac{8}{9m_{Q_{1}}m_{Q_{2}}}\alpha_{s}\left|R_{(ij)}(0)\right|^{2}\label{eq:kappaDq}
\end{eqnarray}
if $Q_{i}$ and $Q_{j}$ are quark and antiquark in color singlet configuration.
Three loop expression for the strong coupling constant was used when calculating
$\kappa_{ij}$ constants \cite{Kiselev:2002iy} and scale for it was taken equal
\begin{eqnarray*}
\mu^{2} & = & \frac{2m_{Q_{1}}m_{Q_{2}}}{m_{Q_{1}}+m_{Q_{2}}}\left\langle T\right\rangle
\end{eqnarray*}
where $\left\langle T\right\rangle $ is average kinetic energy of quarks which is equal
\begin{eqnarray*}
\left\langle T_{d}\right\rangle  & = & 0.19\,\GeV
\end{eqnarray*}
and
\begin{eqnarray*}
\left\langle T_{s}\right\rangle  & = & 0.38\,\GeV
\end{eqnarray*}
for the triplet and singlet states respectively. Values of diquark wave functions in origin
are presented in manuscript \cite{Kiselev:2002iy} while for mesons they can be calculated
from the leptonic width $\Gamma_{ee}$ or leptonic constant $f$ of meson in question:
\begin{eqnarray*}
\left|R(0)\right|^{2} & = & \frac{1}{\alpha^{2}e_{q}^{2}}\Gamma_{ee}\frac{M^{2}}{4}=\frac{M^{2}f}{9}.
\end{eqnarray*}

Up to the hyperfine splitting tetraquark can be described by its total spin $J$,
spins of diquarks $S_{12}$ and $S_{34}$ constituting it and its spatial and charge parities $P$ and $C$:
\begin{eqnarray*}
\left|0^{++}\right\rangle  & = & \left|0;0,0\right\rangle ,\qquad\qquad\qquad\qquad\quad\left|0^{++'}\right\rangle =\left|0;1,1\right\rangle ,\\
\left|1^{+\pm}\right\rangle  & = & \frac{1}{\sqrt{2}}\left(\left|1;0,1\right\rangle \pm\left|1;0,1\right\rangle \right),\qquad\left|1^{+-'}\right\rangle =\left|1;1,1\right\rangle ,\\
\left|2^{++}\right\rangle  & = & \left|2;1,1\right\rangle
\end{eqnarray*}
In this treatment all states are confluent with mass $M_{0}$. If spin-spin interaction is accounted for, masses
of $\left|1^{++}\right\rangle$ and $\left|2^{++}\right\rangle$ states shift:
\begin{eqnarray*}
M\left(1^{++}\right) & = & \left\langle 1^{++}\left|H\right|1^{++}\right\rangle =M_{0}-\kappa_{12}-\kappa_{-},\\
M\left(2^{++}\right) & = & 2m_{[12]}+\kappa_{12}+\kappa_{+},
\end{eqnarray*}
where following designations are introduced:
\begin{eqnarray*}
\kappa_{\pm} & = & \frac{2\kappa_{14}\pm\kappa_{13}\pm\kappa_{24}}{2},
\end{eqnarray*}
and $|0^{++}\rangle$, $|0^{++'}\rangle$ and $|1^{+-}\rangle$,
$|1^{+-'}\rangle$ states mix with each other. In the scalar tetraquarks case
mixing matrix has the following form:
\begin{eqnarray*}
H\left[\begin{array}{c}
\left|0^{++}\right\rangle\\
\left|0^{++'}\right\rangle
\end{array}\right] & = & \left[\begin{array}{cc}
M_{0}-3\kappa_{12} & -\sqrt{3}\kappa_{-}\\
-\sqrt{3}\kappa_{-} & M_{0}+\kappa_{12}-2\kappa_{+}
\end{array}\right]\left[\begin{array}{c}
\left|0^{++}\right\rangle\\
\left|0^{++'}\right\rangle
\end{array}\right],
\end{eqnarray*}
and for $|1^{+-}\rangle$ tetraquarks:
\begin{eqnarray*}
H\left[\begin{array}{c}
|1^{+-}\rangle\\
|1^{+-'}\rangle
\end{array}\right] & = & \left[\begin{array}{cc}
M_{0}-\kappa_{12}+\kappa_{-} & \kappa_{13}-\kappa_{24}\\
\kappa_{13}-\kappa_{24} & M_{0}+\kappa_{12}-\kappa_{+}
\end{array}\right]\left[\begin{array}{c}
|1^{+-}\rangle\\
|1^{+-'}\rangle
\end{array}\right].
\end{eqnarray*}

\subsection{$[QQ][\bar{Q}\bar{Q}]$}

In the case when quarks of the same flavor are involved, Fermi-Dirac statistics
leads to the additional restrictions on the diquark quantum numbers. 
Indeed, permutation of quark indices should change sign of the total diquark wave function.
As quarks are in the antitriplet color state, color part of this function is antisymmetric.
Radial wave function is symmetric as quarks are in the $S$-wave. So spin part of the wave function
is to be symmetric too. Consequently the total spin of the $S$-wave diquark can only be equal to $1$.
As a result only $|0^{++'}\rangle$, $|1^{++'}\rangle$ and $|2^{++}\rangle$
diquark states remain. They do not mix with each other after the spin-spin
interaction is accounted for. Masses of these states are equal
\begin{eqnarray*}
M\left(0^{++'}\right) & = & M_{0}+\kappa_{12}-2\kappa_{+},\\
M\left(1^{+-'}\right) & = & M_{0}+\kappa_{12}-\kappa_{+},\\
M\left(2^{++}\right) & = & M_{0}+\kappa_{12}+\kappa_{+}.
\end{eqnarray*}
It worth mentioning that this splitting scheme agrees with the result of
the work \cite{Berezhnoy:2011xy} in which interaction of the total diquark
spins was concerned.

To obtain numerical values of tetraquark masses one needs to know the unsplitted mass
$M_{0}$ and coefficients $\kappa_{ij}$ in the Hamiltonian (\ref{eq:H}).
These coefficients can be calculated using expressions (\ref{eq:kappaMeson})
and (\ref{eq:kappaDq}). The following values of quark masses were used:
\begin{eqnarray*}
m_{c} & = & 1.468\,\GeV,\qquad m_{b}=4.873\,\GeV.
\end{eqnarray*}
Diquark masses without hyperfine splitting are given in \cite{Kiselev:2002iy}
while $M_{0}$ mass of the tetraquark was calculated using a procedure similar to
that described in \cite{Kiselev:2002iy}.

Let us begin with the tetraquark composed of $4$ $c$-quarks, $T_{4c}=[cc][\bar{c}\bar{c}]$.
Mass of a ground state and value of radial wave function at origin for a
$[cc]$ diquark given in \cite{Kiselev:2002iy} are
\begin{eqnarray*}
m_{[cc]} & = & \mDiqCC\,\GeV,\qquad R_{[cc]}(0)=\RDiqCC\,\GeV^{3/2}.
\end{eqnarray*}
Value of radial wave function at origin of the $(c\bar{c})$ state
determined from the leptonic width of $J/\psi$-meson equals
\begin{eqnarray*}
R_{(c\bar{c})}(0) & = & \RMesonCC\,\GeV^{3/2}.
\end{eqnarray*}
Spin-spin interaction coefficients calculated using expressions
(\ref{eq:kappaMeson}) and (\ref{eq:kappaDq}) are equal
\begin{eqnarray}
\kappa_{12} & = & \kappa_{34}=\kappa_{[cc]}=\kappaDiqCC\,\MeV,\nonumber \\
\kappa_{13} & = & \kappa_{23}=\kappa_{14}=\kappa_{24}=\kappa_{(c\bar{c})}=\kappaMesonCC\,\MeV.\label{eq:kappaCC}
\end{eqnarray}
Without hyperfine splitting $T_{4c}$ tetraquark mass equals
\begin{eqnarray*}
M_{0} & = & \MtetraCC\,\GeV,
\end{eqnarray*}
and with it this state splits into scalar, axial and tensor mesons with masses
\begin{eqnarray*}
0^{++'}: & \qquad & M=\MScalarCC\,\GeV,\qquad M-M_{\mathrm{th}}=\DMScalarCC\,\MeV,\\
1^{+-'}: & \qquad & M=\MAxialCC\,\GeV,\qquad M-M_{\mathrm{th}}=\DMAxialCC\,\MeV,\\
2^{++}: & \qquad & M=\MTensorCC\,\GeV,\qquad M-M_{\mathrm{th}}=\DMTensorCC\,\MeV.
\end{eqnarray*}
In expressions above differences between the tetraquark masses and a $J/\psi$-meson pair
formation threshold are also noted. It can be seen that only tensor state lies
above this threshold and can be observed in the $T_{4c}(2^{++})\to2J/\psi$ mode.
It worth mentioning that scalar tetraquark which is slightly under the $J/\psi$-pair threshold
can however decay by the  $T_{4c}(0^{++'})\to(J/\psi)^{*}J/\psi\to\mu^{+}\mu^{-}J/\psi$ channel i.e.
with one $J/\psi$-meson being virtual. So it can be observed as a peak in the
$\mu^{+}\mu^{-}J/\psi$ invariant mass distribution.

For a tetraquark built from $4$ $b$-quarks, i.e. $T_{4b}=[bb][\bar{b}\bar{b}]$,
situation is entirely similar to the previous case. 
Mass of the $[bb]$ diquark and values of radial wave function at origin
for it and for the $(b\bar{b})$ ground state are
\begin{eqnarray*}
m_{[bb]} & = & \mDiqBB\,\GeV,\qquad R_{[bb]}(0)=\RDiqBB\,\GeV^{3/2},\qquad R_{(b\bar{b})}(0)=\RMesonBB\,\GeV^{3/2}.
\end{eqnarray*}
Mass of the $T_{4b}$ tetraquark without hyperfine splitting is equal to
\begin{eqnarray*}
M_{0} & = & \MtetraBB\,\GeV,
\end{eqnarray*}
and spin-spin interaction coefficients are
\begin{eqnarray}
\kappa_{12} & = & \kappa_{34}=\kappa_{[bb]}=\kappaDiqBB\,\MeV,\nonumber \\
\kappa_{13} & = & \kappa_{23}=\kappa_{14}=\kappa_{(b\bar{b})}=\kappaMesonBB\,\MeV,\label{eq:kappaBB}
\end{eqnarray}
With hyperfine splitting one obtains the following masses of the $T_{4b}$ states:
\begin{eqnarray*}
0^{++'}: & \qquad & M=\MScalarBB\,\GeV,\qquad M-M_{\mathrm{th}}=\DMScalarBB\,\MeV,\\
1^{+-'}: & \qquad & M=\MAxialBB\,\GeV,\qquad M-M_{\mathrm{th}}=\DMAxialBB\,\MeV,\\
2^{++}: & \qquad & M=\MTensorBB\,\GeV,\qquad M-M_{\mathrm{th}}=\DMTensorBB\,\MeV.
\end{eqnarray*}
It can be seen that in this case all the states are under the $\Upsilon(1S)$ pair production
threshold $M_{\mathrm{th}}=2m_{\Upsilon(1S)}$.

\subsection{$[bc][\bar{b}\bar{c}]$}

\begin{figure}
\begin{centering}
\includegraphics[width=12cm]{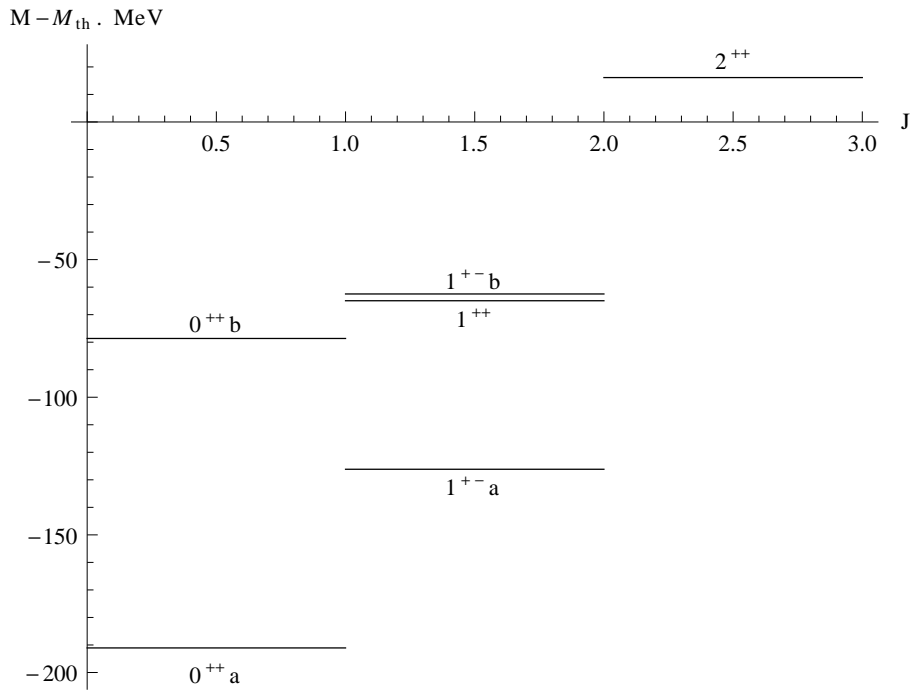}
\par\end{centering}
\caption{$[bc][\bar{b}\bar{c}]$ tetraquark mass spectrum\label{fig:BCmasses}}
\end{figure}

The situation is more interesting in the $T_{2[bc]}=[bc][\bar{b}\bar{c}]$ tetraquark case.
In this case $[bc]$ diquark spin can be $0$ or $1$ and all states mentioned in Section II-A exist.
Diquark mass and value of its radial wave function at origin are \cite{Kiselev:2002iy}
\begin{eqnarray*}
m_{[bc]} & = & \mDiqBC\,\GeV\qquad R_{[bc]}(0)=\RDiqBC\,\GeV^{3/2}.
\end{eqnarray*}
Radial wave function at origin for the color singlet $(b\bar{c})$ state
can be determined by its leptonic constant $f_{B_{c}}=500\,\MeV$ \cite{Kiselev:2002iy}:
\begin{eqnarray*}
R_{(b\bar{c})} & = & \RMesonBC\,\GeV^{3/2}.
\end{eqnarray*}
So spin-spin interaction coefficients are equal
\begin{eqnarray*}
\kappa_{12} & = & \kappa_{34}=\kappa_{[bc]}=\kappaDiqBC\,\MeV,\\
\kappa_{14} & = & \kappa_{23}=\kappa_{(b\bar{c})}=\kappaMesonBC\,\MeV.
\end{eqnarray*}
Values of $\kappa_{13}=\kappa_{(b\bar{b})}$ and $\kappa_{24}=\kappa_{(c\bar{c})}$ constants
were given in expressions (\ref{eq:kappaCC}) and (\ref{eq:kappaBB}). Without hyperfine splitting $T_{2[bc]}$ tetraquark mass equals
\begin{eqnarray*}
M_{0} & = & \MtetraBC\,\GeV,
\end{eqnarray*}
and with spin-spin interaction accounted for this state splits (see Fig.\ref{fig:BCmasses}) into 
\begin{itemize}
\item Two scalar states with masses
\begin{eqnarray*}
0^{++}a: & \qquad & M=\MScalarBCa\,\GeV,\qquad M-M_{\mathrm{th}}=\DMScalarBCa\,\MeV\\
0^{++}b: & \qquad & M=\MScalarBCb\,\GeV,\qquad M-M_{\mathrm{th}}=\DMScalarBCb\,\MeV,
\end{eqnarray*}

\item Two $1^{+-}$ states with masses
\begin{eqnarray*}
1^{+-}a: & \qquad & M=\MAxialBCa\,\GeV,\qquad M-M_{\mathrm{th}}=\DMAxialBCa\,\MeV\\
1^{+-}b: & \qquad & M=\MAxialBCb\,\GeV,\qquad M-M_{\mathrm{th}}=\DMAxialBCb\,\MeV,
\end{eqnarray*}

\item One $1^{++}$ meson with mass
\begin{eqnarray*}
1^{++}: & \qquad & M=\MAxialBCc\,\GeV,\qquad M-M_{\mathrm{th}}=\DMAxialBCc\,\MeV,
\end{eqnarray*}

\item One tensor meson with mass
\begin{eqnarray*}
2^{++}: & \qquad & M=\MTensorBC\,\GeV,\qquad M-M_{\mathrm{th}}=\DMTensorBC\,\MeV.
\end{eqnarray*}

\end{itemize}
Mass of the two $B_c$ mesons is selected for the threshold value in these expressions, $M_{\mathrm{th}}=2m_{B_{c}}=12.55\,\GeV$.
It can be seen that only tensor tetraquark $T_{2[bc]}(2^{++})$ lies above this threshold and thus can be observed
as a peak in the $B_{c}$-meson pair invariant mass distribution.

In paper \cite{Ali:2009pi} tetraquark states were also considered in the framework of diquark model. Picture of hyperfine splittings of $T_{2[bc]}$-tetraquark, presented in this paper is in good agreement with our results. Predictions for masses, on the other hand, are about 700 MeV higher, than our values. As a result, according to this paper all tetraqurk states should lie above $2B_c^*$ and $J/\psi\Upsilon$ thresholds. We think, that the main reason for difference between these two works is the neglection of binding energy in tetraquark tetraquark and diquark spectra, that is negative. For example, for tetraquark state before hyperfine splitting we have $\delta E = M_0 - 2m_{[bc]} \approx 410\,\MeV$.

\section{Production}

\subsection{Duality relations}

Duality relations can be used to estimate production cross sections of
the particles in question. Let us consider formation of two diquarks in
gluon interaction $gg\to[Q_{1}Q_{2}][\bar{Q}_{3}\bar{Q}_{4}]$.
Above the $2$ doubly-heavy baryon production threshold these diquarks
can hadronize into $4$ open heavy flavor mesons, $2$ heavy quarkonia or
form a bound state, i.e. tetraquark. According to our estimations this
tetraquark would preferably decay into a vector quarkonia pair.
Indeed, decay into light mesons is suppressed by the Zweig rule,
production of $4$ open heavy flavor mesons is prohibited kinematically
and formation of pseudoscalar quarkonia requires flip of the heavy quark spin.
That is why the following duality relation can be written:
\begin{eqnarray}
S_{T} & = & \int\limits _{2M_{\mathcal{Q}}}^{2M_{\Xi_{QQ}}}dm_{gg}\hat{\sigma}[gg\to T\to2\mathcal{Q}]=\epsilon\int\limits _{2m_{[QQ]}}^{2M_{\Xi_{QQ}}}dm_{gg}\hat{\sigma}(gg\to[Q_{1}Q_{2}]+[\bar{Q}_{3}\bar{Q_{4}}]),\label{eq:duality}
\end{eqnarray}
where $\epsilon$ factor stands for the other possible decay modes.
This value is to be compared with the integrated non-resonant cross section of
quarkonia pairs production in the same duality window:
\begin{eqnarray}
S_{2\mathcal{Q}} & = & \int\limits _{2M_{\mathcal{Q}}}^{2M_{\Xi_{QQ}}}dm_{gg}\hat{\sigma}\left[gg\to2\mathcal{Q}\right].\label{eq:intCont}
\end{eqnarray}
As tetraquark states are typically narrow, these mesons can be observed as peaks in
the quarkonia pairs invariant mass distributions despite that
$S_{T}\ll S_{2\mathcal{Q}}$ relation holds. As tetraquark width is small compared
to the detector resolution $\Delta\approx50\,\mathrm{MeV}$, its peak
can be modeled by a Gaussian form with corresponding width. Therefore cross section of the
$gg\to T\to2\mathcal{Q}$ Breit-Wigner process is replaced with the following
expression:
\begin{eqnarray*}
\hat{\sigma}(gg\to T\to2\mathcal{Q}) & = & \frac{S_{T}}{\sqrt{\pi}\Delta}\exp\left\{ -\frac{\left(m_{gg}-M_{T}\right)^{2}}{\Delta^{2}}\right\} ,
\end{eqnarray*}
where preexponential factor is selected according to the duality relation (\ref{eq:duality}).

\subsection{$T_{4Q}$}

\begin{figure}
\begin{centering}
\includegraphics[width=8cm]{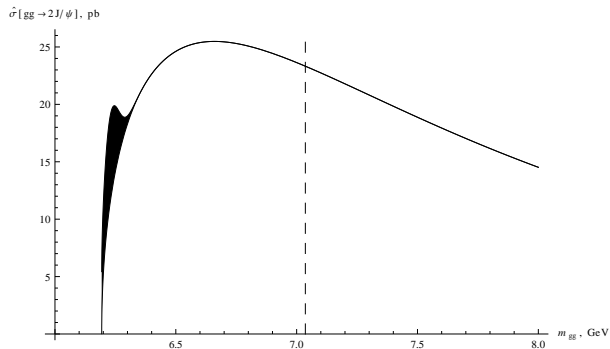}
\par\end{centering}
\caption{Invariant mass distribution of $J/\psi$-meson pairs
with expected $T_{4c}$ tetraquark contribution. Dashed vertical line
corresponds to the two doubly-heavy baryons formation threshold
$2M_{\Xi_{cc}}$.\label{dist:hSigmaX4c}}
\end{figure}

Let us begin with the tetraquarks composed of the $4$ identical quarks.
In the $T_{4c}$ case only tensor state lies above the $2$ vector charmonia
production threshold. Integrated cross sections calculated using expressions (\ref{eq:duality})
and (\ref{eq:intCont}) are equal
\begin{eqnarray*}
S_{T_{4c}} & = & 0.7\,\pb\,\GeV\qquad S_{2J/\psi}=20\ \pb\ \GeV,
\end{eqnarray*}
where suppression factor is selected to be $\epsilon=0.2$. Invariant mass distribution
for the $J/\psi$-meson pairs with expected $T_{4c}$ tetraquark contribution is shown in Fig. \ref{dist:hSigmaX4c}.

As already mentioned, in the $T_{4b}$ tetraquark case even tensor state is under the
two vector bottomonia formation threshold so its observation in their invariant mass distribution
is doubtful.

\subsection{$T_{2[bc]}$}

\begin{figure}
\begin{centering}
\includegraphics[width=16cm]{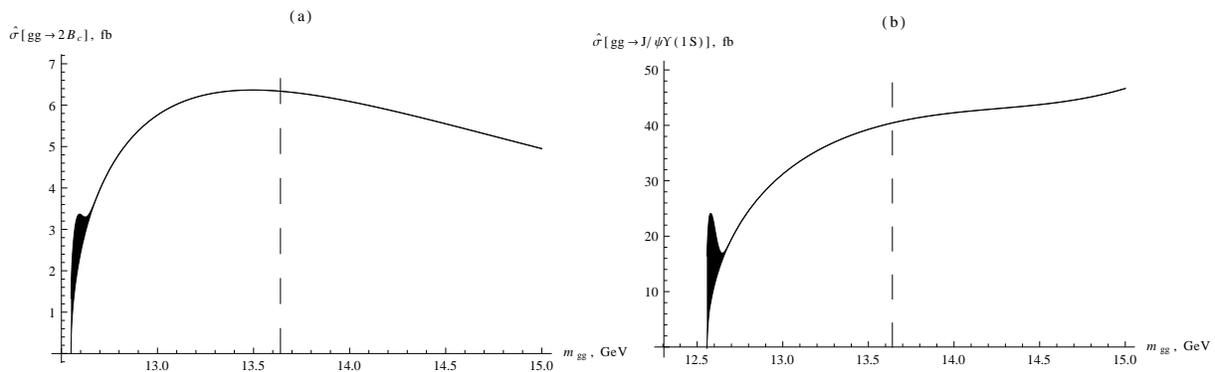}
\par\end{centering}

\caption{Invariant mass distribution of $B_{c}$-meson pairs (left plot)
and $J/\psi\Upsilon(1S)$ (right plot) with expected contribution of the
$T_{2[bc]}$ tetraquark. Dashed vertical line
corresponds to the two doubly-heavy baryons formation threshold $2M_{\Xi_{bc}}$.
\label{dist:hSigma2bc}}
\end{figure}

Let us turn to the $T_{2[bc]}=[bc][\bar{b}\bar{c}]$ tetraquark.
In this case pseudoscalar $B_{c}$-meson and vector $B_{c}^{*}$-meson decaying
into $B_{c}\gamma$ can be experimentally observed.
Thus tensor tetraquark $T_{2[bc]}(2^{++})$ can be observed as a peak in the
$B_{c}$-meson pairs invariant mass distribution. However $T_{2[bc]}(2^{++})\to2B_{c}$ decay requires
flip of the heavy quark spin and is suppressed by the factor
\begin{eqnarray*}
\epsilon & \sim & \frac{M_{T}-2m_{[bc]}}{m_{[bc]}}\approx2.6\times10^{-3}.
\end{eqnarray*}
Integrated cross sections (\ref{eq:duality}) and (\ref{eq:intCont}) are equal
\begin{eqnarray*}
S_{T_{2[bc]}} & = & 0.13\,\fb\,\GeV\qquad S_{2B_{c}}=6\ \fb\,\GeV,
\end{eqnarray*}
where $M_{\Xi_{bc}}=6.82\,\GeV$ \cite{Kiselev:2001fw} is used.
Invariant mass distribution of the $B_c$-meson pairs with expected contribution of the
$T_{2[bc]}$ tetraquark is shown in Fig. \ref{dist:hSigma2bc}a. 
All $T_{2[bc]}$ mesons lie under the $B_{c}^{*}$ pair production threshold.

Decay of the $T_{2[bc]}$ tetraquark into the $J/\psi\Upsilon(1S)$ vector
quarkonia pair is also possible. In the color singlet model reaction
$gg\to J/\psi\Upsilon(1S)$ is prohibited so accounting for octet components
of vector quarkonia is needed. This process was elaborated in the work \cite{Ko:2010xy}
results of which are used as the background for the tetraquark contribution.
Suppression factor $\epsilon\sim3\times10^{-2}$ was used to obtain integrated
cross sections (\ref{eq:duality}) and (\ref{eq:intCont}):
\begin{eqnarray*}
S_{T_{2[bc]}} & = & 1.5\,\fb\,\GeV,\qquad S_{\psi\Upsilon}=33\,\fb\,\GeV.
\end{eqnarray*}
Invariant mass distribution of the $J/\psi\Upsilon(1S)$ pairs
with expected contribution of the $T_{2[bc]}$ tetraquark is shown in Fig. \ref{dist:hSigma2bc}b.

\section{Conclusion}

In our work tetraquarks composed of $4$ heavy quarks are concerned.
Spectroscopy of $T_{4c}=[cc][\bar{c}\bar{c}]$, $T_{4b}=[bb][\bar{b}\bar{b}]$
and $T_{2bc}=[bc][\bar{b}\bar{c}]$ states is elaborated and possibility of 
observation at the LHC is studied.

Model used implies diquark structure of tetraquarks i.e. $Q_{1}Q_{2}\bar{Q}_{3}\bar{Q_{4}}$
tetraquark is assumed to consist of $2$ almost point-like 
diquarks $[Q_{1}Q_{2}]$ and $[\bar{Q}_{3}\bar{Q}_{4}]$ being in triplet color configuration.
In such a treatment tetraquark is entirely analogous to a doubly-heavy meson.
So non-relativistic Schroedinger equation which gives reliable results for the
charmonia and bottomonia description can be used to obtain its parameters.

In the case of $T_{4c}$ and $T_{4b}$ tetraquarks Fermi principle imposes constraints on 
the possible quantum numbers of diquarks: in the antitriplet color configuration 
total spin of the $S$-wave diquark can be only equal to $1$. So with the hyperfine 
splitting accounted for $3$ states with the following quantum numbers appear:
$0^{++}$, $1^{+-}$ and $2^{++}$. In the charmed tetraquark case only tensor meson 
lies above the vector meson pair formation threshold and its peak can be observed at LHC in 
their invariant mass distribution. For the $T_{4b}$ tetraquark all the states lie under
the $\Upsilon(1S)$ pair formation threshold. Both $0$ and $1$ spins are possible for the diquark 
in the $T_{2bc}$ tetraquark. So $6$ states arise after the hyperfine splitting
of the unaffected state: two of $0^{++}$, one $1^{++}$, two of $1^{+-}$ and one $2^{++}$. 

In the last section we estimated the possibility to observe tetraquarks concerned 
in the inclusive reactions $gg\to T\to2\mathcal{Q}$. According to our calculations 
$T_{4c}(2^{++})$ tensor state can be observed as a peak in the $J/\psi$-meson pairs invariant mass distribution.
$T_{2[bc]}(2^{++})$ tetraquark can be observed in both $2B_c$ and $J/\psi\Upsilon(1S)$ modes.

Authors would like to thank A.K. Likhoded and V.V. Kiselev for the fruitful discussions. 
The work was financially supported by Russian Foundation
for Basic Research (grant \#10-02-00061a), non-commercial foundation
``Dynasty'' and the grant of the president of
Russian Federation (grant \#MK-406.2010.2). 


\end{document}